# ANALYSING THE CORRELATION OF GERIATRIC ASSESSMENT SCORES AND ACTIVITY IN SMART HOMES

Björn Friedrich, Enno-Edzard Steen, Sebastian Fudickar and Andreas Hein

Department of Health Services Research, Carl von Ossietzky University, Oldenburg, Germany

## ABSTRACT

*A continuous monitoring of the physical strength and mobility of elderly people is important for maintaining their health and treating diseases at an early stage. However, frequent screenings by physicians are exceeding the logistic capacities. An alternate approach is the automatic and unobtrusive collection of functional measures by ambient sensors. In the current publication, we show the correlation among data of ambient motion sensors and the well-established mobility assessments Short-Physical-Performance-Battery, Tinetti and Timed Up & Go. We use the average number of motion sensor events as activity measure for correlation with the assessment scores. The evaluation on a real-world dataset shows a moderate to strong correlation with the scores of standardised geriatrics physical assessments.*

## KEYWORDS

*ubiquitous computing, biomedical informatics, health, correlation, piecewise linear approximation*

## 1. INTRODUCTION

Being in good health and good physical condition is essential for the quality of life and well-being of humans. Especially, for elderly people who are more prone to diseases and functional decline. Frequently consulting physicians is important for this age group, because early diagnosis is the key for a better treatment and better chances of full recovery. On the one hand, the logistic capacities of physicians are limited and are not sufficient for sophisticated continuous long-term monitoring. On the other hand, long-term monitoring enhances physician's decision-making process. To address this problem unobtrusive smart home sensors can be facilitated for continuous long-term monitoring of elderly people in their domestic environments. Smart home sensors are respecting the privacy of the inhabitant and are well accepted among the target group. They get acquainted to the sensors in a few days and do not notice the sensors anymore [1]. The mobility of elderly people is one key indicator for their physical and mental condition. Moreover, falling is a critical incident for elderly people and even though they recover physically, they may not recover mentally [2-5]. The mobility, balance and muscle-strength of elderly people is usually assessed by physicians or physiotherapists by standardised geriatrics assessments like the Short-Physical-Performance-Battery (SPPB), Timed Up&Go (TUG) and Tinetti test. Those assessments must be performed under the supervision of a professional. Due to capacity issues those assessments cannot be performed frequently. Moreover, the assessment measures the form of the day and people tend to give their best effort in testing situations, in other words there is a difference between performance and capacity. The studies found that the performance is more clinically relevant than the capacity [6].

Our approach uses motion sensor events as indicator for the activity and for the physical conditions of elderly people. We used data from motion sensors installed in domestic environments of elderly people and correlate it with scores of the standardised geriatrics assessments SPPB, Tinetti and TUG. We consider the two parts of the Tinetti separately as Tinetti13 and Tinetti28. Tinetti13 has only balance items and Tinetti28 gait items. This paper is structured as follows:

In Section 2 similar approaches are mentioned and the standardised geriatrics assessments, as state of the art in assessing the physical performance and fall risk in geriatrics, are explained. Section 3 Materials and Methods describes the study for collecting the data, the preparation of the dataset and the used interpolation and correlation methods. In the following result section, the results are explained. In the last section the results are discussed, and an outlook is given.

## 2. STATE OF THE ART

Approved and validated functional tests to assess the physical strength, the mobility and the risk of falling in elderly people are SPPB [7], TUG [8] and Tinetti [9] test. All assessments must be supervised by a professional.

The SPPB assessment has been developed for assessing the mobility of people aged 65 and older. The SPPB assesses the three domains balance, gait speed and strength of the lower limbs. Each domain is assessed by one item and the total performance is scored from 0 to 12 points, where a higher score indicates better mobility and vice versa. The item for assessing the balance is comprised of three sub-items related to balance. The first one is parallel stand, the second is semi-parallel stand and the third one is totally parallel stand. The strength of the lower limbs is assessed by the 5-times Chair Rise item. At the beginning the patient is sitting on a chair and then the patient is asked to stand up and sit down for 5 times in a row without using his or her arms. The gait is assessed by the 4m walk test and the patient is asked to walk over a distance of 4 metres. The time for all assessment items is measured separately and depending on the time the item is scored. The patient can achieve 1 to 4 points for each of the three domains and a total of 12 points. The cut off ranges are 0-6 (low score), 7-9 (middle score), and 10-12 (high score).

The Tinetti test assesses the two domains balance and gait to estimate the risk of falling. The modified version has eight items for balance and another eight for gait. The maximum score for gait performance are 13 points and the maximum score for balance are 15 points. The higher the score, the better the mobility. The items of the Tinetti are on different scales. The balance items are scored from 0 to 4 points, where three items have a score from 0 to 1, four items a score from 0 to 3 and one item from 0 to 4. The gait assessment items are scored from 0 to 2 points and five of the eight items are scored from 0 to 2 and the other three from 0 to 1. The supervisor will score the items in best practice. The scoring depends on the impression of the supervisor because there is a verbal description for giving the points instead of a quantified scale. The cut off scores are 18, 19-23, and 24. A person with a score equal or less than 18 has a high risk of falling, with a score between 19 and 23 a moderate risk of falling, and a score larger or equal 24 a low risk of falling.

The TUG test assesses the mobility of older adults. The score range is from 1 to 4, where 4 is the lowest score and 1 the highest. The test starts with the person sitting on a chair. On the command "Go" the person stands up and walks 3 metres, turns around and walk 3 metres back, and sits down again. The time is measured and based on this measurement the person is scored. The assessment does not only require gait speed, but also both static and dynamic balance. The static balance while sitting and the dynamic balance while standing up and walking. Moreover, the lower limb strength is measured implicitly, because the person must stand up and sit down during the test. A time less than 10s indicates no mobility impairment, 11-19s minor mobility impairment, 20-29s mobility impairment, and greater than 30s severe mobility impairment and intervention is highly recommended.

The approaches to assess the mobility of a person through sensors are, for example, the determination of gait phases and gait parameters, such as step time or length, stride time or length, cadence, gait speed, or maximum toe clearance. These approaches use either wearable or ambient sensors. The wearable sensors are usually inertial sensors, which are positioned at different body locations and detect the movements of one or more parts of the body during walking, are often used as wearable sensors [10]. Typically, inertial sensors are accelerometers, which are used alone or in combination with a triaxial gyroscope, a triaxial magnetometer, or a barometer. Combinations of these sensors are called IMU (Inertial Measurement Unit). An inertial sensor or IMU is used either stand-alone [11-16] or integrated into a smart device such as smartwatch [17] or fitness tracker [18].

Other approaches use pressure or force sensors, either as wearables, e.g. integrated in socks or insoles [19-21] or as ambient sensors, e.g. integrated into sensor carpets [22] or treadmills [23]. Here, the pressure distributions or ground reaction forces are analysed. Besides there is a similar approach that uses capacitive proximity sensors, which can be placed invisibly under different floor coverings and detect the movement of people above [24].

The approaches using video-based systems often determine the positions of joints to detect the movement of the corresponding body parts. These systems can be divided into markerless and marker-based systems. Several markerless approaches use the Microsoft Kinect [25, 26]. Marker-based approaches do not only employ markers, which are placed at anatomically important body positions, e.g. joints, as well as the use of either passive [27] or active markers [28].

Home automation sensors have the advantages of being inexpensive, taking privacy concerns into account, and may already are installed in the domestic environment of a person due to other benefits such as lighting, heating control or security aspects. Typical sensors used to assess the mobility are light barriers [29, 30] and motion sensors. Motion sensors can, for example, are mounted at the ceiling of a frequently used passageway and determine the walking speed of a person [31]. Further approaches analyse the transition times between the coverage areas of different sensors [32-34].

Other sensor-based approaches detect the movements of lower limbs by means of radar [35, 36], laser scanner [37, 38] or ultrasonic sensors [39, 40].

Considering the summary of the state of the art, ambient sensors seem to be the best choice for unobtrusive measurements in domestic environments. Ambient sensors are respecting the privacy and measure the performance and not the capacity, because the person is not engaged in a test situation during the measurements.

## 3. MATERIALS AND METHODS

The used material was a dataset collected during a field study called OTAGO [41]. The main goal of the study was to investigate whether the OTAGO exercise program [47] has an effect in rehabilitation. The used methods are linear approximations for the sensor data and the assessment scores, and a correlation coefficient for the statistical correlation analysis.

## 3.1. Data Acquisition

The data has been collected during the OTAGO study which ranged from July 2014 to December 2015. The planned duration of the study was 40 weeks for each participant. Twenty participants (17 female, 3 male) of an average age of 84.75 years (±5.19 years) participated in the study. They were in pre-frail or frail condition. Due to drop out the average participation time was 36.5 weeks. Due to sickness, visitors, public holidays etc. the average days between two assessments were 31.3 days (±5.3 days). Two participants died during the study and two participants performed the assessments ten times. For the remaining 16 participants eleven assessments have been conducted. At the beginning and every four weeks the standardised geriatrics assessments, Timed Up & Go, SPPB, Barthel Index and Instrumental Activities of Daily Living among others were performed [42-44]. The TUG took longer than 30s for two participants (max. +1.62s). For sake of parity the assessments were scored with 3 points. The characteristics of the study cohort are shown in Table 1 at baseline and Table 2 at the end of the study and the Tables 3 to 14 show the assessment scores of each participant during the study. In addition, ambient passive infrared wireless motion sensors have been installed in the living space of the participants. The motion sensors had a cool down time of 8 seconds when motion cannot be detected. All sensors sent their data over the air to a base station. The sensor system was mainly comprised of home automation sensors and power sensors. A concussion sensor has been placed in the bed, since the used motion sensor was not sensitive enough to measure the small movements while sleeping. A switch with four keys has been installed next to the front door of the homes to indicate whether the person is alone in the flat or not. The participants have been instructed to press a key to make the system aware when another person enters the flat. When the person leaves the flat again or the participant comes home, another key had to be pressed to make the system aware that only one person is inside the flat. In Figure 1 a flat of one of the participants is shown.

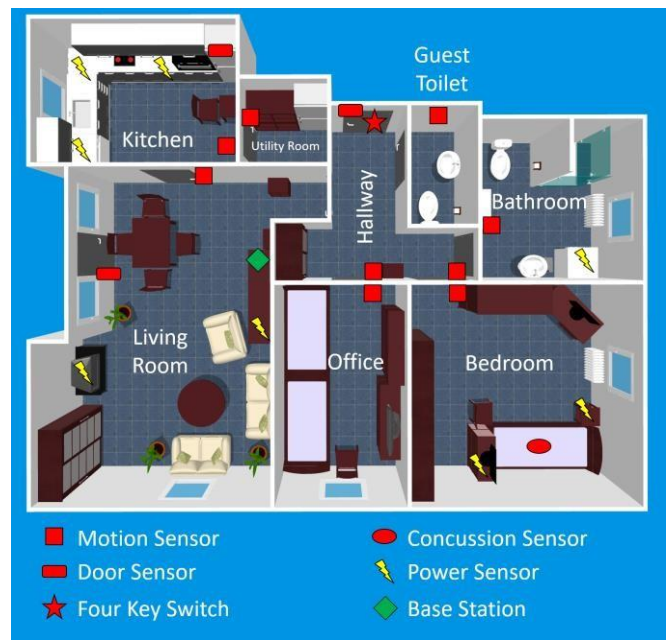

Figure 1. Example of a flat of one of the participants. The positions of the sensors are marked by symbols.

Table 1. The baseline characteristics of the study cohort.

| n=20, m=3,f=17 | Age (y) | Frailty Index(pts) | iADL (pts) | SPPB (pts) | Tinetti (pts) | TUG (s) |
|---|---|---|---|---|---|---|
| Mean | 84.8 | 1.9 | 7.3 | 6.0 | 22.5 | 17.9 |
| SD | 5.2 | 0.7 | 1.4 | 2.3 | 4.7 | 5.3 |
| Range (min - max) | 76.0 – 92.0 | 1.0 – 3.0 | 3.0 – 8.0 | 3.0 – 11.0 | 13.0 – 28.0 | 11.2 – 31.6 |

Table 2. The characteristics of the study cohort at the end (T10)

| n=18, m=3, f=15 | Age (y) | Frailty Index (pts) | iADL (pts) | SPPB (pts) | Tinetti (pts) | TUG (s) |
|---|---|---|---|---|---|---|
| Mean | 84.5 | 2.0 | 6.1 | 6.6 | 20.9 | 16.4 |
| SD | 4.9 | 1.0 | 2.3 | 2.9 | 5.7 | 6.0 |
| Range (min –max) | 77.0 –93.0 | 0.0 –4.0 | 1.0 –8.0 | 2.0 –12.0 | 7.0 –27.0 | 8.5 –30.1 |

Table 3. The assessment scores of participant 1. The first row is the month of the study. All values are in points and the TUG scores are seconds (points).

| ID 1 | 1 | 2 | 3 | 4 | 5 | 6 | 7 | 8 | 9 | 10 | 11 |
|---|---|---|---|---|---|---|---|---|---|---|---|
| SPPB | 3 | 3 | 3 | 3 | 3 | 2 | 3 | 2 | 3 | 4 | 4 |
| Tinetti13 | 5 | 7 | 11 | 5 | 7 | 9 | 8 | 7 | 6 | 8 | 8 |
| Tinetti28 | 14 | 18 | 20 | 12 | 15 | 18 | 16 | 14 | 13 | 15 | 14 |
| TUG | 31.6 (3) | 27.2 (3) | 24.2 (3) | 28.9 (3) | 26.6 (3) | 25.0 (3) | 22.1 (3) | 24.7 (3) | 21.9 (3) | 21.7 (3) | 22.3 (3) |

Table 4. The assessment scores of participant 2. The first row is the month of the study. All values are in points and the TUG scores are seconds (points).

| ID 2 | 1 | 2 | 3 | 4 | 5 | 6 | 7 | 8 | 9 | 10 | 11 |
|---|---|---|---|---|---|---|---|---|---|---|---|
| SPPB | 8 | 7 | 7 | 5 | 8 | 9 | 8 | 10 | 7 | 8 | 7 |
| Tinetti13 | 11 | 11 | 13 | 12 | 13 | 12 | 11 | 13 | 13 | 13 | 12 |
| Tinetti28 | 25 | 25 | 28 | 26 | 26 | 25 | 23 | 26 | 27 | 28 | 23 |
| TUG | 14.5 (2) | 15.7 (2) | 13.6 (2) | 13.1 (2) | 13.3 (2) | 11.7 (2) | 13.0 (2) | 11.8 (2) | 13.4 (2) | 16.2 (2) | 13.1 (2) |

Table 5. The assessment scores of participant 3. The first row is the month of the study. All values are in points and the TUG scores are seconds (points).

| ID 3 | 1 | 2 | 3 | 4 | 5 | 6 | 7 | 8 | 9 | 10 | 11 |
|---|---|---|---|---|---|---|---|---|---|---|---|
| SPPB | 8 | 8 | 11 | 11 | 10 | 10 | 10 | 11 | 11 | 12 | 10 |
| Tinetti13 | 11 | 13 | 13 | 13 | 13 | 12 | 13 | 13 | 12 | 12 | 12 |
| Tinetti28 | 24 | 27 | 27 | 26 | 26 | 26 | 27 | 25 | 26 | 25 | 26 |
| TUG | 14.3 (2) | 11.1 (2) | 10.0 (1) | 10.8 (2) | 10.1 (2) | 10.5 (2) | 7.9 (1) | 8.4 (1) | 7.7 (1) | 7.7 (1) | 8.5 (1) |

Table 6. The assessment scores of participant 4. The first row is the month of the study. All values are in points and the TUG scores are seconds (points).

| ID 4 | 1 | 2 | 3 | 4 | 5 | 6 | 7 | 8 | 9 | 10 | 11 |
|---|---|---|---|---|---|---|---|---|---|---|---|
| SPPB | 9 | 9 | 11 | 9 | 10 | 9 | 10 | 10 | 7 | 9 | 11 |
| Tinetti13 | 13 | 13 | 13 | 13 | 12 | 12 | 12 | 13 | 12 | 13 | 13 |
| Tinetti28 | 28 | 28 | 28 | 26 | 25 | 25 | 25 | 28 | 26 | 26 | 26 |
| TUG | 17.6 (2) | 11.3 (2) | 16.7 (2) | 12.8 (2) | 12.8 (2) | 12.4 (2) | 12.9 (2) | 11.2 (2) | 15.7 (2) | 12.3 (2) | 11.9 (2) |

Table 7. The assessment scores of participant 5. The first row is the month of the study. All values are in points and the TUG scores are seconds (points).

| ID 5 | 1 | 2 | 3 | 4 | 5 | 6 | 7 | 8 | 9 | 10 | 11 |
|---|---|---|---|---|---|---|---|---|---|---|---|
| SPPB | 5 | 10 | 8 | 7 | 7 | 8 | 8 | 7 | 8 | 7 | 8 |
| Tinetti13 | 11 | 11 | 12 | 13 | 13 | 13 | 13 | 13 | 13 | 12 | 12 |
| Tinetti28 | 23 | 26 | 27 | 28 | 28 | 27 | 28 | 28 | 27 | 27 | 27 |
| TUG | 14.0 (2) | 10.2 (2) | 11.9 (2) | 11.3 (2) | 10.8 (2) | 9.2 (1) | 11.3 (2) | 10.8 (2) | 10.9 (2) | 12.1 (2) | 11.8 (2) |

Table 8. The assessment scores of participant 6. The first row is the month of the study. All values are in points and the TUG scores are seconds (points). Due to medical condition the assessment scores of month 7 are not available.

| ID 6 | 1 | 2 | 3 | 4 | 5 | 6 | 7 | 8 | 9 | 10 | 11 |
|---|---|---|---|---|---|---|---|---|---|---|---|
| SPPB | 4 | 3 | 4 | 3 | 4 | 3 | N/A | 4 | 4 | 4 | 4 |
| Tinetti13 | 8 | 8 | 8 | 8 | 6 | 6 | N/A | 8 | 7 | 10 | 7 |
| Tinetti28 | 18 | 21 | 17 | 18 | 15 | 13 | N/A | 17 | 15 | 17 | 13 |
| TUG | 23.5 (3) | 15.8 (2) | 20.7 (3) | 19.0 (3) | 22.9 (3) | 20.4 (3) | N/A | 19.2 (2) | 21.0 (3) | 19.9 (3) | 21.6 (3) |

Table 9. The assessment scores of participant 7. The first row is the month of the study. All values are in points and the TUG scores are seconds (points).

| ID 7 | 1 | 2 | 3 | 4 | 5 | 6 | 7 | 8 | 9 | 10 | 11 |
|---|---|---|---|---|---|---|---|---|---|---|---|
| SPPB | 4 | 4 | 5 | 4 | 3 | 4 | 5 | 5 | 4 | 5 | 5 |
| Tinetti13 | 6 | 6 | 8 | 9 | 7 | 7 | 6 | 8 | 8 | 10 | 10 |
| Tinetti28 | 16 | 15 | 19 | 19 | 16 | 16 | 15 | 17 | 18 | 20 | 20 |
| TUG | 21.0 (3) | 16.8 (2) | 15.8 (2) | 15.5 (2) | 17.7 (2) | 16.0 (2) | 14.6 (2) | 14.9 (2) | 14.1 (2) | 15.5 (2) | 15.5 (2) |

Table 10. The assessment scores of participant 8. The first row is the month of the study. All values are in points and the TUG scores are seconds (points). The participant deceased after participating 8 months.

| ID 8 | 1 | 2 | 3 | 4 | 5 | 6 | 7 | 8 | 9 | 10 | 11 |
|---|---|---|---|---|---|---|---|---|---|---|---|
| SPPB | 6 | 10 | 9 | 11 | 10 | 9 | 9 | 11 | N/A | N/A | N/A |
| Tinetti13 | 13 | 13 | 12 | 13 | 12 | 13 | 12 | 11 | N/A | N/A | N/A |
| Tinetti28 | 27 | 26 | 26 | 26 | 25 | 27 | 24 | 26 | N/A | N/A | N/A |
| TUG | 12.0 (2) | 11.1 (2) | 12.8 (2) | 12.9 (2) | 13.3 (2) | 12.7 (2) | 10.4 (2) | 9.9 (1) | N/A | N/A | N/A |

Table 11. The assessment scores of participant 9. The first column is the row of the study. All values are in points and the TUG scores are seconds (points). The participant deceased after participating 3 months.

| ID 9 | 1 | 2 | 3 | 4 | 5 | 6 | 7 | 8 | 9 | 10 | 11 |
|---|---|---|---|---|---|---|---|---|---|---|---|
| SPPB | 4 | 10 | 4 | N/A | N/A | N/A | N/A | N/A | N/A | N/A | N/A |
| Tinetti13 | 12 | 11 | 8 | N/A | N/A | N/A | N/A | N/A | N/A | N/A | N/A |
| Tinetti28 | 25 | 26 | 17 | N/A | N/A | N/A | N/A | N/A | N/A | N/A | N/A |
| TUG | 19.5 (2) | 20.1 (2) | 23.4 (3) | N/A | N/A | N/A | N/A | N/A | N/A | N/A | N/A |

Table 12. The assessment scores of participant 10. The first row is the month of the study. All values are in points and the TUG scores are seconds (points). Due to medical condition the assessment scores of month 6 are not available.

| ID 10 | 1 | 2 | 3 | 4 | 5 | 6 | 7 | 8 | 9 | 10 | 11 |
|---|---|---|---|---|---|---|---|---|---|---|---|
| SPPB | 3 | 3 | 2 | 3 | 2 | N/A | 2 | 2 | 2 | 2 | 2 |
| Tinetti13 | 7 | 6 | 6 | 5 | 6 | N/A | 3 | 3 | 1 | 2 | 4 |
| Tinetti28 | 13 | 15 | 12 | 12 | 11 | N/A | 6 | 6 | 5 | 5 | 7 |
| TUG | 21.3 (3) | 25.4 (3) | 27.2 (3) | 22.4 (3) | 21.3 (3) | N/A | 24.8 (3) | 24.4 (3) | 23.9 (3) | 28.7 (3) | 30.1 (3) |

Table 13. The assessment scores of participant 11. The first row is the month of the study. All values are in points and the TUG scores are seconds (points).

| ID 11 | 1 | 2 | 3 | 4 | 5 | 6 | 7 | 8 | 9 | 10 | 11 |
|---|---|---|---|---|---|---|---|---|---|---|---|
| SPPB | 5 | 3 | 3 | 7 | 5 | 8 | 7 | 5 | 6 | 5 | 5 |
| Tinetti13 | 10 | 9 | 5 | 9 | 8 | 6 | 9 | 10 | 11 | 9 | 10 |
| Tinetti28 | 23 | 16 | 13 | 19 | 16 | 15 | 18 | 17 | 18 | 16 | 18 |
| TUG | 13.8 (2) | 19.3 (2) | 23.8 (3) | 15.8 (2) | 14.3 (2) | 15.3 (2) | 11.8 (2) | 15.7 (2) | 12.8 (2) | 19.6 (3) | 17.5 (2) |

Table 14. The assessment scores of participant 12. The first row is the month of the study. All values are in points and the TUG scores are seconds (points).

| ID 12 | 1 | 2 | 3 | 4 | 5 | 6 | 7 | 8 | 9 | 10 | 11 |
|---|---|---|---|---|---|---|---|---|---|---|---|
| SPPB | 4 | 5 | 5 | 6 | 3 | 6 | 3 | 2 | 3 | 4 | 4 |
| Tinetti13 | 6 | 13 | 10 | 11 | 10 | 11 | 10 | 10 | 11 | 11 | 11 |
| Tinetti28 | 17 | 27 | 20 | 20 | 19 | 18 | 18 | 18 | 22 | 21 | 19 |
| TUG | 23.1 (3) | 19.3 (2) | 19.0 (2) | 20.3 (3) | 21.5 (3) | 21.8 (3) | 24.2 (3) | 26.4 (3) | 19.6 (3) | 24.3 (3) | 22.4 (3) |

### 3.2. Preprocessing

The data described in Section 3.1. is preprocessed in the following manner. The sensor events of each day are added up for each sensor. Then the average number of events per day is computed by adding up the number of events and dividing it by the number of motion sensor in the flat of the participant. The result is one feature per day. The mathematical formulation is

$$\frac{1}{n} \sum_{i=1}^{n} \sum_{j=1}^{10800} \mathbf{1}_A(e_{i,j}) \tag{1}$$

Where *n* is the number of sensors installed in the flat, j the 8 seconds time window of the day and 1 the indicator function defined on the set A of all sensor events. In other words, the indicator function is 1 if there is a sensor event e from sensor i in time window j and 0 if there is no event. If there is no sensor event recorded on a certain day, the day is excluded from the dataset. The average days between two assessments after removing are 31.6 days and the assessment scores were used as they were recorded. Unless a sub-item could not be performed, but the remaining items can, the sub-item is scored with 0 even though the items score cannot be 0 according to the manual.

Several participants have been excluded from the dataset. Three participants were excluded, because they were hospitalised during the study. Their data was incomplete and after being discharged from the hospital the participants used walking frames and got assistance while performing the assessments. In three flats the motion sensors in key areas have been installed a few months after the study started. Hence, the data from the most frequently used rooms like the kitchen, living room and hallway is not available. These three participants have been excluded as well. Another two participants have been excluded due to incomplete data, there was an error that caused fragmented data. Overall, we excluded eight participants from the analysis. Such exclusion resulted in a final cohort of 12, with 10 female and 2 male participants.

## 3.3. Interpolation and Approximation

Two different interpolation methods were used for the values. The assessment scores are interpolated using a spline interpolation and the average activities per day are approximated with a linear regression. The piecewise polynomial interpolation or spline interpolation is an ordinary linear function defined as follows

$$s(x) = m \cdot x + b \tag{2}$$

where *x* is the date of the assessment, *m* the slope and *b* the interception with the y-axis. In addition, for each two consecutive scores $a_i$ and $a_{i+1}$ the following conditions must hold

$$\begin{aligned} a_i &= s(x_i) = m_i \cdot x_i + b_i \\ a_{i+1} &= s(x_{i+1}) = m_i \cdot x_{i+1} + b_i \end{aligned} \tag{3}$$

where *i* denotes the index of the assessment score. Spline interpolation is used, because the assessments were taken in an average interval of 31.3 days and assuming a linear change is feasible. The frequency of the average motion in one day is much higher. Between two assessments an average of 31.6 values are available. This value is slightly larger than the average days between two assessments, because we excluded some participants from our dataset. Linear regression is more robust in the face of outliers than spline interpolation. So, linear regression is used to approximate a function for the average motion values. The linear regression has the same base function as the spline interpolation, but the way of computing the values *m* and *b* is different

$$\arg\min_m \sum_{i=1}^n d(m \cdot x_i, v_i) \tag{4}$$

where *d* is an arbitrary metric function, *i* the number of values and $v_i$ the *i*-th value of the value set. Formula IV is computed for different *m*'s and the *m* which results in the smallest sum is chosen as best parameter for the regression. For this research, the Euclidean distance is used as metric. The linear regression formula is not taking *b* into account. However, after computing *m* there is only one unknown left in the equation. Using linear algebra, the unique solution can be computed.

The interpolated and fitted values are correlated with each other using Spearman's ρ.

### 3.4. Correlation Coefficient and Thresholds

For correlation, the Spearman Rank Correlation or Spearman's ρ is used [45]. The correlation assesses whether there is a monotonic relationship between two variables. In contrast to the Pearson Correlation there is only one assumption that must hold. It is sufficient when the variables are in an ordinal scale. To each value its rank is assigned. The values are sorted in an ascending order and the rank is the index of the value. Since, two values can have the same rank, the rank is not well-defined. To overcome this, the equal values are slightly altered to become different and the new rank is the mean of the ranks of the altered values. This is called Ties. Once all ranks are assigned the correlation is computed with the formula

$$\rho = \frac{\sum_{i=0}^{n}(R(x_i) - \mu(R_x))(R(y_i) - \mu(R_y))}{\sqrt{\sum_{i=0}^{n}(R(x_i) - \mu(R_x))^2}\sqrt{\sum_{i=0}^{n}(R(y_i) - \mu(R_y))^2}} \quad (5)$$

where R($x_i$) denotes the rank of value $x_i$, $\mu$ the mean of all ranks of the corresponding variable and *n* is the number of values. The formula reveals if all values for one variable are equal, the correlation coefficient is not defined, because a division by zero occurs.

For judging the strength of the correlation, the definition of Cohen [46] is used. Correlations between *0.1* and *0.3* are considered as small, between *0.3* and *0.5* are considered as moderate and larger than *0.5* are considered as large. This holds for the negative values as well. A correlation is statistically significant when *p<0.001* holds.

## 4. RESULTS

All correlations satisfying the threshold of *0.3* are significant with a *p*-Value smaller than 0.001.

All participants have at least one assessment with a moderate correlation. The smallest correlation is *0.3* for participant *2* with the SPPB and with the Tinetti13 assessments. The smallest significant correlation is the correlation with the SPPB of participant *10* with *0.23*. The *p*-Values of each smaller correlation is greater than 0.001. Participant *9* has the largest correlation values over all for all assessments. There is only one participant (3) with one assessment with a correlation stronger than moderate. The participants *1,2,6,11* have a correlation stronger than *0.3* for two assessments. The Tinetti13 and Tinetti28 are correlated for participant *11*. The SPPB and Tinetti13 are correlated only for participant 2 and SPPB and Tinetti28 are correlated only for participant *1*. For participants 8 all assessments except for the Tinetti28 are correlated with a minimum of *0.43*. For participant *9* all assessments are correlated with a minimum correlation of *0.82*. The TUG assessment is the only assessment where sometimes the correlation could not be computed. The scores of the participants 1,2,4 and 10 were not changing during the study. The largest correlation with *0.88* is found for participant *9* and the smallest significant correlation for participant *12*. For participants 5,8,9, and 12 the magnitude of the correlation of TUG and SPPB are similar, e.g. for participant 5 both correlations are moderate. All the participants, where the TUG correlation is not applicable, are showing small changes in all assessment scores and the scores are never crossing a cut off score. The correlation values and corresponding *p*-values are shown in Table 15.

Table 15. The participants and the correlations with the assessments SPPB, Tinetti13, Tinetti28 and TUG. Correlations that are moderate at least are in bold font; N/A means Not Applicable.

| ID | Assessment Correlation | | | |
|---|---|---|---|---|
| | **SPPB** | **Tinetti13** | **Tinetti28** | **TUG** |
| 1 | **-0.56** (p<0.001) | 0.10 (p<0.07) | **0.43** (p<0.001) | N/A (N/A) |
| 2 | **-0.30** (p<0.001) | **0.30** (p<0.001) | 0.26 (p<0.001) | N/A (N/A) |
| 3 | -0.12 (p<0.02) | **-0.32** (p<0.001) | -0.01 (p<0.8) | -0.04 (p<0.5) |
| 4 | **-0.50** (p<0.001) | -0.15 (p<0.006) | **-0.42** (p<0.001) | N/A (N/A) |
| 5 | **-0.33** (p<0.001) | 0.14 (p<0.01) | -0.20 (p<0.001) | **0.37** (p<0.001) |
| 6 | 0.03 (p<0.6) | **-0.40** (p<0.001) | **-0.46** (p<0.001) | 0.01 (p<0.92) |
| 7 | -0.05 (p<0.3) | **0.70** (p<0.001) | 0.28 (p<0.001) | 0.24 (p<0.001) |
| 8 | **0.49** (p<0.001) | **-0.43** (p<0.001) | -0.10 (p<0.001) | **-0.52** (p<0.001) |
| 9 | **0.88** (p<0.001) | **0.82** (p<0.001) | **0.84** (p<0.001) | **-0.88** (p<0.001) |
| 10 | 0.23 (p<0.001) | **0.60** (p<0.001) | -0.25 (p<0.001) | N/A (N/A) |
| 11 | -0.06 (p<0.2) | **-0.61** (p<0.001) | 0.26 (p<0.001) | **0.47** (p<0.001) |
| 12 | **-0.34** (p<0.001) | 0.02 (p<0.7) | **-0.37** (p<0.001) | **0.33** (p<0.001) |

Correlating the scores achieved in the three domains of the SPPB leads to the results shown in Table 16. A moderate to large correlation is found for the participants *2,3,4,6,8* and *12*. Participant *9* has a large positive correlation for all three domains. The domain balance correlates with the average motion sensor events for the participants *6,8,9*, the domain gait and 4 metres correlate for the participants *2,4,9*, and the domain assessing the strength of the lower limbs correlates for participants *3,9*, and *12*. There is no moderate correlation found for participants *1,5,7,10*, and *11*. There is a correlation of 0.0 for participant 1 with 5CRT, participant *7* for 5CRT as well and for participant *10* for balance and 4 metres.

Table 16. The correlation of the three domains assessed by the SPPB. 5 times chair rise and 4m gait test. Correlations that are moderate at least are in bold font.

| ID | SPPB Item Correlation | | |
|---|---|---|---|
| | **Balance** | **4m** | **5CRT** |
| 1 | -0.22 (p<0.001) | -0.21 (p<0.001) | 0.00 (p<0.0) |
| 2 | 0.01 (p<0.7) | **-0.63** (p<0.001) | -0.21 (p<0.001) |
| 3 | -0.10 (p<0.05) | 0.26 (p<0.001) | **0.36** (p<0.001) |
| 4 | 0.20 (p<0.001) | **-0.62** (p<0.001) | -0.23 (p<0.001) |
| 5 | -0.25 (p<0.001) | -0.20 (p<0.001) | -0.22 (p<0.001) |
| 6 | **-0.58** (p<0.001) | 0.17 (p<0.007) | -0.13 (p<0.04) |
| 7 | -0.01 (p<0.83) | -0.15 (p<0.008) | 0.00 (p<0.0) |
| 8 | **0.52** (p<0.001) | -0.21 (p<0.002) | 0.14 (p<0.04) |
| 9 | **0.82** (p<0.001) | **0.82** (p<0.001) | **0.82** (p<0.001) |
| 10 | 0.00 (p<0.0) | 0.00 (p<0.0) | -0.06 (p<0.2) |
| 11 | 0.13 (p<0.01) | -0.02 (p<0.6) | 0.26 (p<0.001) |
| 12 | -0.25 (p<0.001) | 0.24 (p<0.001) | **-0.78** (p<0.001) |

# 5. DISCUSSION

Table 15 shows some interesting findings. Even though the three assessments are assessing similar domains the correlations are different. The reason why SPPB and Tinetti13 have a different correlation is that Tinetti13 is comprised of balance items only while the SPPB includes additional parameters that cover gait and lower limb muscle-strength. A good example for such variational effect is participant 12. For this participant, the correlation of SPPB is moderate and there is no correlation with Tinetti13 and a weak correlation with Tinetti28. Looking at Table 2 the gait which is assessed in Tinetti13 and the balance which is assessed in Tinetti13 and Tinetti28 shows a weak correlation only, but the 5CRT which is assessing the lower limb strength has a large correlation. The structure of the SPPB and the TUG is the reason for the similar correlations. Both assessments are assessing the same dimensions. In the SPPB the three dimensions are divided into three different items, the TUG assesses the three dimensions indirectly. The balance and the lower limb strength are assessed by standing up and sitting down, and the gait by the 3m walk.

With participant 9 showing the highest correlation overall, the general validity of the ambient motion sensors to detect functional decline can be confirmed. It is worthwhile, to investigate the individual history of this case: Two month in the study, the participant got a chemotherapy therapy. Therefore, the physical and psychological conditions of this participant became worse rapidly. Due to the frequent treatments in the hospital amount of data is small compared to the other participants. While the corresponding decline is well given in this case, others slighter trajectories as well have been present.

For participant 2 there is moderate correlation for SPPB and Tinetti13. The explanation is that only the 4m gait test has large correlation and the other two items have no and a weak correlation respectively. The SPPB takes all three domains into account equally and the Tinetti13 is comprised of items for assessing gait only. The Tinetti28 is slightly imbalanced towards the balance items because the maximum balance score is higher than the maximum gait score.

Even considering Table 16 there is no explanation for some combination of correlations. For participants 1,5,7,10, and 11 there is no significant correlation for the SPPB items, but there are moderate to strong correlations found for the assessments themselves. The reason might be a combination of the items of the assessments. To verify this further investigation is needed.

The unclear results could be traced back to the study as well. The sensors were installed in the domestic environments and could not be controlled. Some sensors were relocated by the dweller so that the sensing area changed. That might led to a blind spot, where a lot of activities were done. That would have changed the number of events and the cause is not a change in mobility, but in sensor relocation.

From the medical point of view the participants with the strongest mobility impairment show stronger or more correlations. The correlation is at least large (>0.6) for one assessment or moderate for two or more assessments. The reason is that their condition is very volatile, and the assessment results are only representing the form of the day. Moreover, the participants in good condition can show good performance in the assessments, but may are sluggish during none assessment times. The correlation of the SPPB of participant 8 is caused by the strong improvements in mobility. Participant 8 is the only participant, except for participant 9, where the SPPB scores are improving from the lowest to the highest cut off interval. The variety of the correlations and the combination of correlating assessments reflects the variety of the study cohort.

Even though there are correlations found, it is too early to conclude causalities or medical reasons. Further medical studies are needed to closely investigate the relation between the motion sensor events and the assessment scores.

## 6. CONCLUSION AND FUTURE WORK

The results show that the approach using motion sensor data for assessing the mobility of elderly people is feasible for continuous long-term monitoring and provides valuable information for physicians. The correlations found with SPPB and Tinetti are moderate (≥0.3) at least and statistically significant ($p<0.001$).

There are two ways to further investigate the relation between the motion sensor data and the assessment scores. The first way is to improve the interpolation, regression and analytical methods. Artificial intelligence algorithms show promising results in ubiquitous computing and analysing data from distributed sensor systems. So, the second way is to add more information to the data and additional data from other sensors. Power consumption sensors can add valuable information about activities for further analysis. The current data does not take the entropy of a sensor event into account. For example, a motion sensor which is attached near the door to the backyard might not have as many events as a motion sensor in the living room, but the information that the participant left the flat is more important than the participant is in the living room. Moreover, the sequence of the events could be taken into account. Those sequences can give information about the ways of the participant in the flat. The ratio between unnecessary ways in the flat and necessary ones like going to the toilet, may proof to be a good feature to improve the correlation.

In addition, there are unclear correlation combinations, maybe due to special combinations of assessment items might be the cause. To find an explanation the correlations of the items must be explored further by correlating every single Tinetti item with the average motion sensor events.

### ACKNOWLEDGEMENTS

We acknowledge Prof. Dr. Jürgen Bauer (University of Heidelberg) for designing and supervising the OTAGO study. We acknowledge Bianca Sahlmann (University of Oldenburg) and Lena Elgert (Peter L. Reichertz Institute, Hannover) for performing the assessments. The OTAGO study has been funded by an internal funding of the Carl von Ossietzky University of Oldenburg and has been approved by the ethics committee under the ethics vote Drs.72/2014.

### REFERENCES

[1] Marschollek, M. & Becker, M. & Bauer, J. & Bente, P. & Elgert, L. & Elbers, K. & Hein, A. & Kolb, G. & Künemund, H. & Lammel-Polchau, C. & Meis, M. & Schwabedissen, H. & Remmers, H. & Schulze, M. & Steen, E.-E. & Thoben, W. & Wang, J. & Wolf, K.-H. & Haux, R., (2014) „Multimodal activity monitoring for home rehabilitation of geriatric fracture patients
– feasibility and acceptance of sensor systems in the GAL-NATARS study", Informatics for Health and Social Care, Vol. 39, pp262-271

[2] Phillips, L. J. & DeRoche, C. B. & Rantz, M. & Alexander, G. L. & Skubc, M. & Despins, L. & Abbot, C. & Harris, B. H. & Galambos, C. & Koopman, R. J., (2017) "Using embedded sensors in independent living to predict gait changesand falls", Western Journal of Nursing Research, Vol. 39, No. 1, pp78–94.

[3] Studenski, S. & Perera, S. & Patel, K. & Rosano, C. & Faulkner, K. & Inzitari, M. & Brach, J. & Chandler, J. & Cawthon, P. & Connor, E. B. & Nevitt, M. & Visser, M. & Kritchevsky, S. &

Badinelli, S. & Harris, T. & Newman, A. B. & Cauley, J. & Ferrucci, L. & Guralnik, J., (2011) "Gait speed and survival in older adults", JAMA, Vol. 305, No. 1, pp50–58.
[4] Middleton, A. & Fritz, S. J. & Lusardi, M., (2015) "Walking speed: the functional vital sign", Journal of Aging and Physical Activity, Vol. 23, No. 2, pp314–322.
[5] Shuman, V. & Coyle, P. C. & Perera, S. & VanSwearingen, J. M. & Albert, S. M. & Brach, J. S., (2020) "Association between improved mobility and distal health outcomes", The Journals of Gerontology Series A Biological Sciences and Medical Sciences.
[6] Giannouli, E. & Bock, O. & Mellone, S. & Zijlstra, W., (1994) "Mobility in old age: Capacity is not performance", BioMed Research International, Vol. 2016.
[7] Guralnik, J. M. & Simonsick, E. M. & Ferrucci, L. & Glynn, R. J. & Berkman, L. F. & Blazer, D. G. & Scherr, P. A. & Wallace, R. B., (1994) "A short physical performance battery assessing lower extremity function: Association with self-reported disability and prediction of mortality and nursing home admission", Journal of Gerontology, Vol. 49, ppM85–M94.
[8] Podsiadlo D. & Richardson, S., (1991) "The Timed Up & Go: A test of basic functional mobility for frail elderly persons", Journal of the American Geriatrics Society, Vol. 32, pp142–148.
[9] Tinetti, M. E., (1986) "Performance-oriented assessment of mobility problems in elderly patients", Journal of the American Geriatrics Society, Vol. 34, pp119–126.
[10] Sprager, S. & Juric, M. B., (2015) "Inertial sensor-based gait recognition: A review ", Sensors (Basel, Switzerland), Vol. 15, pp22089–22127.
[11] Moon, Y. & McGinnis, R. S. & Seagers, K. & Motl, R. W. & Sheth, N.& Wright, J. A. & Ghaffari, R. & Sosnoff, J. J., (2017) "Monitoring gait in multiplesclerosis with novel wearable motion sensors", PloS one, Vol. 12, p.e0171346.
[12] Raccagni, C. & Gaßner, H. & Eschlboeck, S. & Boesch, S. & Krismer, F. & Seppi, K. & Poewe, W. & Eskofier, B. M. & Winkler, J. & Wenning, G. & Klucken, J., (2018) "Sensor-based gait analysis in atypical parkinsonian disorders", Brain and behavior, Vol. 8, pe00977.
[13] Schlachetzki, J. C. M. & Barth, J. & Marxreiter, F. & Gossler, J. & Kohl, Z. & Reinfelder, S. & Gassner, H. & Aminian, K. & Eskofier, B. M. & Winkler, J. & Klucken, J., (2017) "Wearable sensors objectively measure gait parameters in parkinson's disease", PloS one, Vol. 12, pe0183989.
[14] Terrier, P. & Le Carre, J. & Connaissa, M.-L. & Leger, B. & Luthi, F., (2017) "Monitoring of gait quality in patients with chronic pain of lower limbs", IEEE transactions on neural systems and rehabilitation engineering : a publication of the IEEE Engineering in Medicine and Biology Society, Vol. 25, pp1843–1852.
[15] Manor, B. & Yu, W. & Zhu, H. & Harrison, R. & Lo, O.-Y.& Lipsitz, L. & Travison, T. & Pascual-Leone, A. & Zhou, J., (2018) "Smartphone app–based assessment of gait during normal and dual-task walking: demonstration of validity and reliability", JMIR mHealth and uHealth, Vol. 6, No. 1, pe36.
[16] Jung, D. & Nguyen, M. D. & Park, M. & Kim, M. & Won, C. W. & Kim, J. & Mun, K.-R., (2020), „ Walking-in-Place Characteristics-Based Geriatric Assessment Using Deep Convolutional Neural Networks ", 42nd Annual International Conference of the IEEE Engineering in Medicine & Biology Society (EMBC), Montreal, Canada, pp. 3931-3935.
[17] Erdem, N. S. & Ersoy, C. & Tunca, C., (2019) "Gait analysis using smart-watches", Proc. Indoor and Mobile Radio Communications (PIMRC Workshops) IEEE 30th Int. Symp. Personal, pp1– 6.
[18] Floegel, T. A. & Florez-Pregonero, A. & Hekler, E. B. & Buman, M. P., (2017) "Validation of consumer-based hip and wrist activity monitors in older adults with varied ambulatory abilities", The journals of gerontology Series A Biological sciences and medical sciences, Vol. 72, pp229– 236.
[19] Tirosh, O. & Begg, R. & Passmore, E. & Knopp-Steinberg, N., (2013) "Wearable textile sensor sock for gait analysis", Proc. Seventh Int. Conf. Sensing Technology (ICST), pp618–622.
[20] Saidani, S. & Haddad, R. & Mezghani, N. & Bouallegue, R., (2018) "A survey on smart shoe insole systems", International Conference on Smart Communications and Networking (SmartNets) IEEE, pp1–6.
[21] Luna-Perejón, F. & Domíngues-Morales, M. & Guitiérrez-Galán, D. & Civit-Balcells, A., (2020), "Low-Power Embedded System for Gait Classification Using Neural Networks", Journal of Low Power Electronics and Applications (Sensors), Vol. 10, No. 2:14.
[22] CIR Systems (USA), (2020) "GAITRite®walkways" https://www.gaitrite.com/gait-analysis-walkways, online; last accessed: 2020-02-23.
[23] Bertec Corporation (USA), (2020) "Instrumented treadmills", https://www.bertec.com/products/instrumented-treadmills, online; last accessed: 2020-02-23.

[24] Future-Shape GmbH (Germany), (2020) "Sensfloor", https://future-shape.com/en/system, online; last accessed: 2020-02-23.
[25] Dubois, A. & Bresciani, J.-P., (2018) "Validation of an ambient system for the measurement of gait parameters", Journal of biomechanics, Vol. 69, pp175–180.
[26] Springer, S. & Yogev Seligmann, G., (2016) "Validity of the kinect for gait assessment: A focused review", Sensors (Basel, Switzerland), Vol. 16, p194.
[27] Vicon Motion Systems Ltd. (UK), (2020) "Vicon nexus", https://www.vicon.com/software/nexus, online; last accessed: 2020-02-23.
[28] Northern Digital Inc. (Canada), (2020) "Optotrak certus", https://www.ndigital.com/msci/products/optotrak-certus, online; last accessed: 2020-02-23.
[29] Frenken, T. & Steen, E.-E. & Brell, M. & Nebel, W. & Hein, A., (2011) "Motion pattern generation and recognition for mobility assessments in domestic environments", AAL 2011 - Proceedings of the 1st International Living Usability Lab Workshop on AAL Latest Solutions, Trends and Applications, pp3–12.
[30] Hein, A. & Steen, E.-E. & Thiel, A. & Hülsken-Giesler, M. & Wist, T. & Helmer, A. & Frenken, T. & Isken, M. & Schulze, G. C. & Remmers, H., (2014) "Working with a domestic assessment system to estimate the need of support and care of elderly and disabled persons: results from field studies", Informatics for Health and Social Care, Vol. 39, No. 3-4, pp210–231.
[31] Hagler, S. & Austin, D. & Hayes, T. L. & Kaye, J. & Pavel, M., (2010) "Unobtrusive and ubiquitous in-home monitoring: A methodology for continuous assessment of gait velocity in elders", IEEE transactions on biomedical engineering, Vol. 57, No. 4, pp813–820.
[32] Aicha, A. N. & Englebienne, G. & Kröse, B., (2017) "Continuous measuring of the indoor walking speed of older adults living alone", Journal of ambient intelligence and humanized computing, pp1–11.
[33] Hellmers, S. & Steen, E.-E. & Dasenbrock, L. & Heinks, A. & Bauer, J. M. & Fudickar, S. & Hein, A., (2017) "Towards a minimized unsupervised technical assessment of physical performance in domestic environments", Proceedings of the 11th EAI International Conference on Pervasive Computing Technologies for Healthcare, pp207–216.
[34] Rana, R. & Austin, D. & Jacobs, P. G. & Karunanithi, M. & Kaye, J., (2017) "Gait velocity estimation using time-interleaved between consecutive passive ir sensor activations", IEEE Sensors Journal, Vol. 16, No. 16, pp6351–6358.
[35] Rui, L. & Chen, S. & Ho, K. C. & Rantz, M. & Skubic, M., (2017) "Estimation of human walking speed by doppler radar for elderly care", JAISE, Vol. 9, No. 2, pp181–191.
[36] Wang, F. & Skubic, M. & Rantz, M. & Cuddihy, P. E., (2014) "Quantitative gait measurement with pulse-Doppler radar for passive in-home gait assessment", IEEE Transactions on Biomedical Engineering, Vol. 61, No. 9, pp2434–2443.
[37] Fudickar, S. & Stolle, C. & Volkening, N. & Hein, A., (2018) "Scanning laser rangefinders for the unobtrusive monitoring of gait parameters in unsupervised settings", Sensors (Basel, Switzerland), Vol. 18.
[38] Iwai, M. & Koyama, S. & Tanabe, S. & Osawa, S. & Takeda, K. & Motoya, I. & Sakurai, H. & Kanada, Y. & Kawamura, N., (2019) "The validity of spatiotemporal gait analysis using dual laser range sensors: a cross-sectional study", Archives of physiotherapy, Vol. 9, p3.
[39] Ferre, X. & Villalba-Mora, E. & Caballero-Mora, M.-A. & Sanchez, A. & Aguilera, W. & Garcia-Grossocordon, N. & Nunez-Jimenez, L. & Rodriguez-Manas, L. & Liu, Q. & del Pozo Guerrero, F., (2017) "Gait speed measurement for elderly patients with risk of frailty", Mobile Information Systems, Vol. 2017, p11.
[40] Qi, Y. & Soh, C. B. & Gunawan, E. & Low, K.-S. & Thomas, R., (2016) "Assessment of foot trajectory for human gait phase detection using wireless ultrasonic sensor network", IEEE transactions on neural systems and rehabilitation engineering: a publication of the IEEE Engineering in Medicine and Biology Society, Vol. 24, pp88–97.
[41] Carl von Ossietzky University, (2020). "OTAGO", https://uol.de/en/amt/research/projects/otago, online; last accessed: 2021-01-08
[42] Mahoney, F. & Barthel, D., (1965) "Functional evaluation: The Barthel index", Maryland State Medical Journal, Vol. 14, pp56–61.
[43] Searle, S. D. & Mitnitski, A. & Gahbauer, E. A. & Gill, T. M. & Rockwood, K., (2008) "A standard procedure for creating a frailty index", BMC Geriatrics, Vol. 8.

[44] Lawton, M. P. & Brody, E. M., (1969) "Assessment of older people: Self-maintaining and instrumental activities of daily living", The Gerontologist, Vol. 9, pp179–186.
[45] Spearman, C., (1904) "The proof and measurement of association between two things", The American Journal of Psychology, Vol. 15, No. 1, pp72–101.
[46] Cohen, J., (1988) "Statistical Power Analysis for the Behavioral Sciences", Lawrence Erlbaum Associates.
[47] Campbell, A. J. & Robertson, C., (2010) "Comprehensive Approach to Fall Prevention on a National Level: New Zealand", Clinics in Geriatric Medicine, Vol. 26, No. 4, pp719-731